\documentclass[sigconf]{acmart}

\usepackage{subcaption}
\usepackage{amsmath}
\usepackage{pifont}  
\usepackage{graphicx}
\usepackage{multirow}

\AtBeginDocument{%
  }

\setcopyright{acmlicensed}
\copyrightyear{2018}
\acmYear{2018}
\acmDOI{XXXXXXX.XXXXXXX}
\acmConference[Conference acronym]{Conference Name}{Date}{Venue}
\acmISBN{978-1-4503-XXXX-X/2018/06}

\begin{document}

\title{MRG-Bench: Evaluating and Exploring the Requirements of Context for Repository-Level Code Generation}

\author{Haiyang Li}
\affiliation{%
  \institution{Peking University}
  \city{Beijing}
  \country{China}
}
\author{Qing Gao}
\affiliation{%
  \institution{Peking University}
  \city{Beijing}
  \country{China}
}
\author{Shikun Zhang}
\affiliation{%
  \institution{Peking University}
  \city{Beijing}
  \country{China}
}

\begin{abstract}
Large Language Models (LLMs) have demonstrated impressive capabilities in code generation. However, current evaluation datasets suffer from issues such as the lack of runnable test cases, deviation from the distribution of real-world code, and the ability to evaluate only the Python language. These limitations undermine the credibility of the evaluation results.

To address these limitations, we introduce \textbf{MRG-Bench} (Multi-language Repository-level Code Generation Benchmark), a novel dataset that provides a more accurate evaluation of LLMs in practical repository-level code generation tasks. MRG-Bench has three main features: (1) practical data sourced from real-world code repositories that align to the practical distribution, (2) multiple programming languages support, including Python, Java, and Go, and (3) project-level runnable test cases to assess the quality of the generated code.

Based on MRG-Bench, we conducted extensive experiments including large language models, long-context models, and RAG-related methods. These evaluation results demonstrate that \textbf{current repository-level code generation techniques suffer from significant performance deficiencies}. To further investigate why models fail, we designed novel experiments to annotate the underlying causes of generation errors. The results explicitly show that the majority of methods suffer from "\textbf{difficulty in understanding user requirements}," failing to comprehend their assigned tasks accurately. Moreover, the impact of different repository-level contexts on this issue exhibits significant disparities across different programming languages, suggesting that, in practice, specialized contextual information needs to be designed for different languages. 

\end{abstract}



\received{20 February 2007}
\received[revised]{12 March 2009}
\received[accepted]{5 June 2009}

\maketitle

\begin{figure}[h]
    \centering
    \begin{subfigure}[b]{1\linewidth}
        \includegraphics[width=\linewidth]{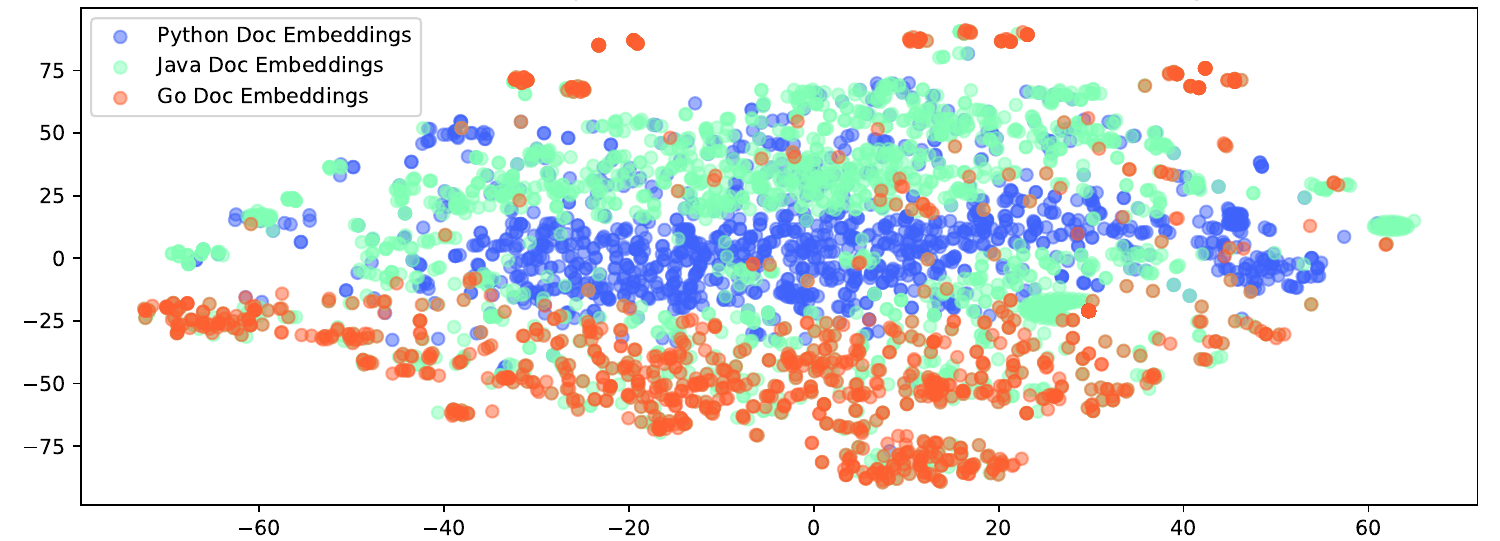}
        \caption{The semantic distributions of function comments across three different programming languages exhibit significant differences, highlighting the importance of evaluating each language independently.}
        \label{fig:comb_a}
    \end{subfigure}
    
    \vspace{0.5cm}
    
    \begin{subfigure}[b]{1\linewidth}
       \includegraphics[width=\linewidth]{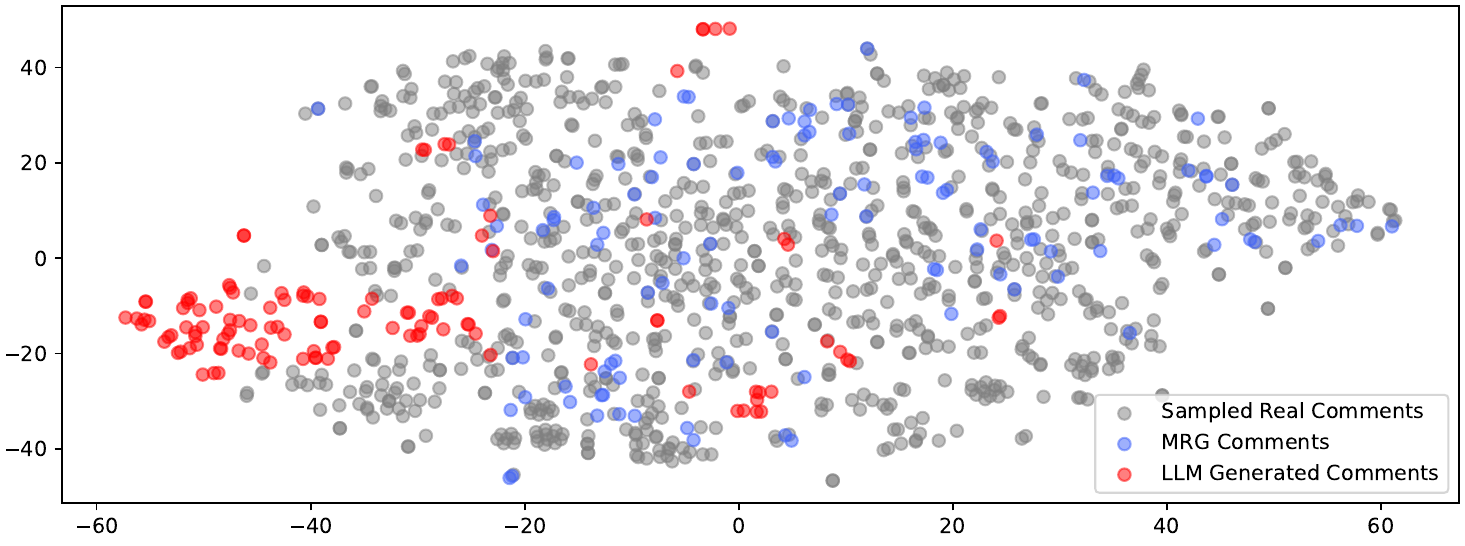}
        \caption{The semantic distributions of function comments of real-world data, LLM-generated data, and MRG-bench.}
        \label{fig:comb_b}
    \end{subfigure}
\end{figure}

\begin{table*}
  \centering
  \caption{Comparison of different datasets}
  \label{tab:dataset_comparison}

  \begin{tabular}{lcccc}
    \hline
    \textbf{dataset} & \textbf{multi-language support} & \textbf{repository information} & \textbf{project runnable environment} & \textbf{practical data source} \\
    \hline
    NumpyEval~\cite{numpyeval} & \ding{55} & \checkmark & \ding{55} & \ding{55} \\
    ClassEval~\cite{classeval} & \ding{55} & \ding{55} & \ding{55} & \ding{55} \\
    CoderEval~\cite{codereval} & \checkmark & \ding{55} & \ding{55} & \checkmark \\
    EvolCodeBench~\cite{evobench} & \ding{55} & \checkmark & \checkmark & \ding{55} \\
    AgentBench~\cite{codeagent} & \ding{55} & \checkmark & \checkmark & \ding{55} \\
    SWEBench~\cite{swebench} & \ding{55} & \checkmark & \checkmark & \checkmark \\
    RepoCoder~\cite{repocoder} & \ding{55} & \checkmark & \ding{55} & \checkmark \\
    \hline
    MRG-Bench & \checkmark & \checkmark & \checkmark & \checkmark \\
    \hline
  \end{tabular}
\end{table*}

\section{Introduction}
\label{sec:intro}
The remarkable performance of large language models (LLMs) in code generation tasks has garnered significant attention in recent years, with related research steadily emerging~\cite{codellama,deepseekcoder}. The introduction of commercial products like GitHub Copilot~\cite{github} further highlights the practical potential of LLMs. These models not only assist developers in generating code but also enhance efficiency in various aspects of the development process. However, current evaluation methods \cite{codereval,coderujb,codexglue,classeval} for LLMs in code generation predominantly focus on standalone code fragments, which limits their applicability in real-world development scenarios. In practice, development is typically conducted within independent code repositories, involving complex code dependencies, modular architectures, and multiple programming languages. Thus, evaluating the code generation capabilities of LLMs within the context of an actual code repository has become a key focus in ongoing research.

Several efforts \cite{repocoder,repobench,evobench} have proposed repository-level code generation datasets, but they exhibit the following limitations: (1) Limited programming languages: These datasets predominantly focus on specific programming languages, such as Python. However, as shown in Figure~\ref{fig:comb_a}, the distributions of semantic embeddings from three different programming languages exhibit distinct differences. In fact, different programming languages often excel in specific domains of code. Therefore, a multi-language evaluation dataset can better assess the performance of LLMs (Large Language Models) in real-world programming scenarios.  (2) Inconsistent with the practical situation: Some work \cite{evobench} attempts to use large language models to generate code summaries, treating these summaries as natural language descriptions of the code.  As shown in Figure~\ref{fig:comb_b}, there is a significant semantic discrepancy between the requirement descriptions generated by LLMs and the actual code requirements. (3) Lack of a runnable environment: Although some approaches generate code snippets that appear highly similar to the target code, these snippets may fail to compile or pass tests when executed in a real-world environment. (4) Limited Evaluation Scope: Previous studies have predominantly attempted to extract fixed contexts from repositories and evaluate large language models accordingly. However, in practice, different retrieval methods are often employed to recall relevant code snippets from repositories, or more powerful reasoning models are utilized for generation. The absence of evaluation for these methods in prior work has impacted the accurate assessment of current technologies. These three limitations undermine the reliability of previous evaluations of repository-level code generation tasks. 

Addressing the problems above, we first propose a new evaluation dataset, MRG-Bench(Multi-language repository level code Generation Benchmark), designed to more accurately reflect the real-world performance of LLMs in repository-level code generation. Subsequently, we conducted extensive evaluation experiments based on this dataset, encompassing multiple language models, long-context models, reasoning models, as well as common RAG methods. Furthermore, we attempted to design a novel data annotation dimension to investigate why current methods perform poorly on MRG-Bench.

For the dataset, MRG-Bench has three key features: \textbf{(1) Practical data source:} All data in the dataset is collected from real-world code repositories, ensuring that the code snippets reflect the actual distribution of code in real development scenarios. \textbf{(2) Multi-language coverage:} In addition to Python (the most common language), MRG-Bench includes other programming languages, such as Java and Go, enabling a more comprehensive evaluation of LLMs’ code generation capabilities in different languages. \textbf{(3) Runnable test cases:} Each project in the dataset includes executable test cases, ensuring that the code generated by the model can be run in real-world environments, thereby enhancing the credibility of the evaluation results.  As shown in Table~\ref{tab:dataset_comparison}, current datasets exhibit significant gaps across critical features, whereas MRG-Bench demonstrates more comprehensive assessment capabilities. It is noteworthy that although we include SWE-Bench~\cite{swebench} in the table, its target scenario differs fundamentally from MRG-Bench; SWE-Bench is designed for bug fixing according to real issues rather than code generation scenarios.
 
For the evaluation, we conduct an extensive evaluation of state-of-the-art models, including large language models, reasoning models, long-context models, and several RAG-based methods.  Our experiments highlight the following key findings: (1) \textbf{Poor performance of LLMs on MRG-Bench}: Current large language models perform inadequately on the MRG-Bench dataset. Even the best language model, Claude-3.5-Sonnet, achieves an average Pass@1 score of only 32.5\%, indicating significant room for improvement. (2) \textbf{Language bias in large language models}: LLMs exhibit a noticeable bias towards specific programming languages. All models perform better in Python when relevant contextual information is lacking. (3) \textbf{Ineffectiveness of RAG-related methods}: Retrieval-Augmented Generation (RAG) methods perform poorly, even worse than simply providing the file information containing the function. 

Based on the findings before, we further investigated the underlying causes of failures in code generation across different contexts. Specifically, we decomposed the model generation process into two perspectives: (1) understanding user requirements (knowing \textbf{What} to do) and (2) implementing user requirements (knowing \textbf{How} to do it). We employed model-based annotation for failed cases. The results reveal several interesting findings: \textbf{(1) The majority of failure cases occur in "What to do", where models fail to comprehend what users intend to generate. (2) Different repository contexts exhibit pronounced disparities in their effects on performance.} For instance, intra-file content significantly aids model comprehension of user requirements in Java and Go, while showing negligible impact in Python. These two experimental findings indicate that current models are most struggling at requirement understanding. Moreover, methods for enhancing this stage vary across different programming languages. Consequently, in real-world engineering applications, we may need to dynamically adjust supplementary contextual content based on the specific programming language to achieve superior performance.

In conclusion, our work makes the following contributions:

1. \textbf{Practical dataset with multiple languages:} We published a multi-language dataset, MRG-Bench, with a real-world data distribution and runnable test cases, offering a more realistic evaluation benchmark for future studies. Besides, we released the function call graph analysis framework which enables researchers to quickly generate private evaluation datasets for their own research.
 
2. \textbf{Comprehensive evaluation and analysis:} We performed an extensive evaluation of the performance of popular large language models on MRG-Bench, along with an analysis of the basic RAG methods. Our results demonstrate that current models need improvement when applied to real-world tasks.

3. \textbf{Novel Analytical Methodology}: We designed a novel analytical approach to investigate the underlying causes of failures and to examine the influence of varying contextual information on the results. Our experimental findings demonstrate that the primary cause of failures lies in the inability to accurately comprehend user inputs, thereby indicating that enhancements targeting this particular perspective can rapidly improve the final performance.

We hope this work provides valuable insights into the application of large language models in real-world coding scenarios and inspires future research.

\begin{figure}
    \centering
    \includegraphics[width=1\linewidth]{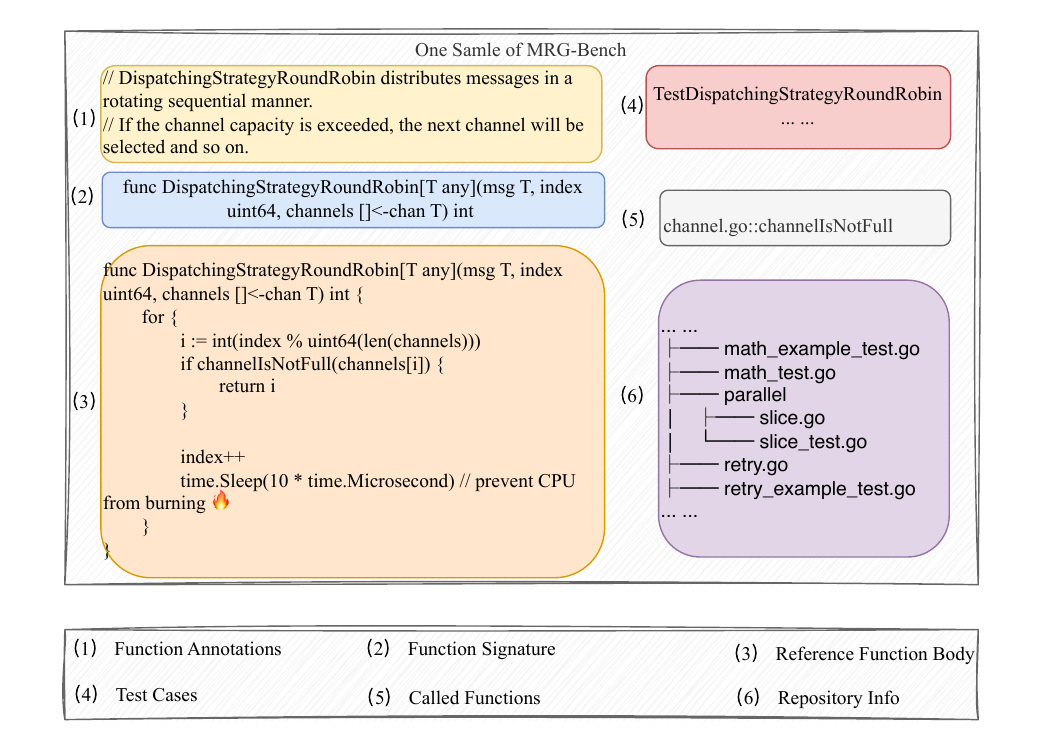}
    \caption{All Information in One Sample of MRG-Bench}
    \label{fig:sample}
\end{figure}

\section{MRG-Bench Description}

\textbf{MRG-Bench} includes three programming languages and contains a total of 383 samples. Each sample is a function-level code generation task which includes the following information, as shown in Figure~\ref{fig:sample}:

\begin{enumerate}
    \item Function annotations: Natural language comment used for generating the function.
    \item Function signature: function name, parameters, and return type (if any).
    \item Referenced Function Body: code snippet from repository.
    \item Test Cases: unit test or integration test cases for this sample.
    \item Called Private Functions: A list of other functions in the repository that the function calls.
    \item Repository Information: The original repository.
\end{enumerate}

\begin{table}
\centering
\caption{Language Statistics}

\begin{tabular}{ccccc}
\hline
\textbf{Language} & \textbf{Samples} & \textbf{Avg Query Tokens} & \textbf{Avg Target Tokens}\\ 
\hline
Python & 124 & 80.90 & 283.00\\ 
Java & 96 & 70 & 152.83\\ 
Go & 163 & 38.36 & 153.23\\ 
\hline
\end{tabular}
\label{tab:language_stats}
\end{table}

\section{MRG-Bench Construction}

Our dataset construction adheres to three primary principles:

\textbf{Practical:} The dataset is derived from real-world open-source projects, avoiding artificially constructed data. The function signatures, natural language descriptions of code functions, and test cases are all sourced from actual projects. Besides, the distribution of queries should closely approximate the distribution of queries in real-world code repositories.

\textbf{Multi-Language:} The dataset should include different programming languages to facilitate comprehensive evaluations.

\textbf{Executability:} The ultimate evaluation criterion for the dataset is that the code generated by models should work in real-world scenarios, passing both unit and integration tests in the project.

The construction of our dataset is divided into following key steps. 

\subsection{Programming language selection}

We select three languages in this version of MRG-Bench: Python, Go, and Java, which have robust package management approaches and unit test frameworks. The primary reason for excluding C and C++ was the difficulty of managing conflicting dependencies in constructing runnable environments in Linux Docker containers (planning for next version, see Section~\ref{sec:threats}). These projects often depend on specific Linux distributions and various dynamic libraries, for which there is no efficient way to manage environments. Although JavaScript has a robust package management system that allows for quick configuration of its runtime environment, we found that the code comment ratio and unit test coverage were low. Additionally, most test suites were integration tests, making it challenging to match each test case to its corresponding function through automated analysis. Therefore, we finally selected Python, Java, and Go as the supported languages of our datasets.

\subsection{Selection of code repositories}

For the selected programming languages, we utilized the GitHub API to retrieve repositories. To ensure that the large language model (LLM) has limited prior knowledge of the repositories, we focused on repositories created after January 2023, following previous work~\cite{repocoder}. We ranked these repositories by the number of stars and manually reviewed repositories over 50 stars, following previous work~\cite{evobench}. The repositories were filtered based on the following criteria:

\begin{enumerate}
    \item The repository should include a complete test suite, containing at least 10 unit test functions.
    \item The repository must be able to run on the Linux platform.
    \item The repository should compile and execute successfully, passing its most test cases.
\end{enumerate}

Based on these criteria, we finally selected 152 repositories from a total of 1,000 repositories across 3 languages.

\subsection{Function dependency construction}

After obtaining a project that can be compiled and tested locally, we extracted the functions required for our dataset. These functions were selected based on the following criteria:
\begin{enumerate}
    \item They have dependencies on other functions within the repository.
    \item They include developer-written function comments.
    \item They have corresponding test cases within the project.
\end{enumerate}

To extract functions that meet these requirements, we developed project-level function call graph analyzers for each language. Specifically, we first utilized Tree-sitter to parse the source code into function definitions and class method definitions. For each function, we analyzed the files (import info and package info) and symbols referenced (variable defined info) within the function, and matched function calls to project-defined functions. This process allowed us to construct a function call graph for all custom functions in the project.

Next, we identified test functions by checking whether the function name, class name, file name, or path contained the keyword "test." Each test function was then linked to the function(s) it called, marking it as a test case for those functions. Finally, we saved all non-test functions that contain documentation and corresponding test cases as candidate data. In this phase, we further filtered out repositories that did not yield valid data (i.e., those that did not produce functions containing both test cases and function annotations), ultimately leaving 23 projects, 580 samples remaining.

\subsection{Manual inspection and Test Coverage Filtering}

In the final step of dataset construction, we reviewed the obtained dataset to ensure that each function includes meaningful annotations. We identified and filtered out annotations that do not provide meaningful descriptions of the function’s behavior, such as generic comments like "return String" or "implement interface ClientInterface." Additionally, we manually executed the unit tests of these projects and utilized corresponding tools (i.e., \texttt{pytest-cov}, \texttt{Jacoco}, and \texttt{go test -cover}) to generate function-level test coverage information. We then removed all functions that did not achieve 100\% line coverage. Ultimately, we obtained a total of 383 samples from 22 projects.

\subsection{Evaluation Method and Metrics}

We have configured the runtime environment and test command for each project in a docker container. When executing the tests, we copy the function to be tested to the target position of the file in the project, and then run the corresponding test case for that sample. Specifically, for Python projects, we use \texttt{pytest} to execute the test cases. For Java projects, we utilize Maven to manage dependencies and run unit tests using the \texttt{mvn test} command. For Go projects, we use \texttt{go test} to execute the specified test cases. 

All projects are configured within an executable Linux container, where the necessary environments have already been set up. Given the complexity and time required to build this container from scratch, we do not recommend that researchers attempt to do so. Instead, you can pull our pre-configured Docker image from Docker Hub to quickly verify your method locally.

We use the \texttt{Pass@k} metric as the performance evaluation metric for MRG-Bench. \texttt{Pass@n} represents the proportion of samples which has at least one generation result that passes all test cases.

\section{Experiment Setting}

\subsection{Model Selection}

We selected six large language models of varying sizes for experiments. These models have demonstrated strong coding capabilities and achieved competitive scores on the HumanEval dataset~\cite{humaneval}. Open-source models include DeepSeek-Coder-33B~\cite{deepseekcoder}, CodeLLaMA-13B~\cite{codellama}, LLaMA-3.1-8B-Instruct\footnote{https://github.com/meta-llama/llama-models}, and StarChat2-15B~\cite{starcoder2}, while the closed-source models are Claude3.5-Sonnet \footnote{https://claude.ai/} and GPT-4o \footnote{https://chatgpt.com/}. For reasoning models, we select o3-mini\footnote{https://chatgpt.com/}, Deepseek-R1~\cite{deepseekr1}, and Qwen2.5-QwQ-32B~\cite{qwq}.

\subsection{Experimental Setup}

For the open-source models, we deployed them locally using vLLM \cite{vllm}. For the closed-source models, we accessed GPT-4o via the Azure API\footnote{https://azure.microsoft.com/} and Claude3.5-Sonnet through VertexAI's API \footnote{https://console.cloud.google.com/vertex-ai/}. We set the temperature to 0.6 to balance the stability and creativity of the model's generated outputs, while the remaining parameters were configured using either the API defaults or the default settings provided by vLLM.  Due to the request frequency limitations of closed-source models, we could not include them in all experiments. For the subsequent RAG-based experiments, we utilized DeepSeek-Coder-33B, as it performed best in the previous evaluation in open-source LLMs. The RAG experiments were conducted using the LangChain\footnote{https://www.langchain.com/} framework. All the code for our experiments is publicly available on GitHub\footnote{https://github.com/MRG-Bench/MRG-Bench}.

\subsection{Research Questions}

In this paper, we conduct experiments to answer the following research questions:

\textbf{RQ1: Why do we need MRG-Bench?}
    Corresponding to the motivations for constructing MRG-Bench outlined in Section \ref{sec:intro}, we designed experiments to demonstrate the advantages of MRG-Bench over existing datasets from two perspectives. 

\textbf{RQ1.1 Is MRG-Bench more representative compared to existing repository-level code generation datasets?}
    We analyzed the semantic distribution of samples in MRG-Bench to show that our dataset aligns more closely with real-world code distributions compared to existing datasets. Consequently, results obtained on this dataset better reflect the true performance of models. 
    
\textbf{RQ1.2 Is Multi-language evaluation necessary for current LLMs?}
    We conducted experiments on mainstream large language models to empirically demonstrate the necessity of a multi-language dataset for evaluating current models, thereby highlighting the advantage of MRG-Bench’s multi-language support.
    
\textbf{RQ2: How do current LLMs perform on MRG-Bench with different contexts?}

    Lacking information about the function to be generated will decrease the performance of the models. In this question, we explore the impact of different contextual information on the model. In this question, we primarily designed two types of experimental settings: static context and dynamic context. The static context encompasses 2 configurations: in-file context and available utility function information. In the dynamic context experiments, we evaluated multiple RAG-related methods, including basic RAG approaches based on BM25 retrieval and embedding retrieval methods, as well as RepoCoder, an algorithm specifically designed for repository-level code generation.

\textbf{RQ3: Why do current methods perform poorly on MRG-Bench?}

    In this question, we attempt to annotate the failure cases and analyze why the models fail on these samples, as well as examine the impact of different contexts on the model's failure cases.

\section{Result and analysis}

\begin{figure}
    \centering
    \begin{subfigure}[b]{0.75\linewidth}
        \includegraphics[width=\linewidth]{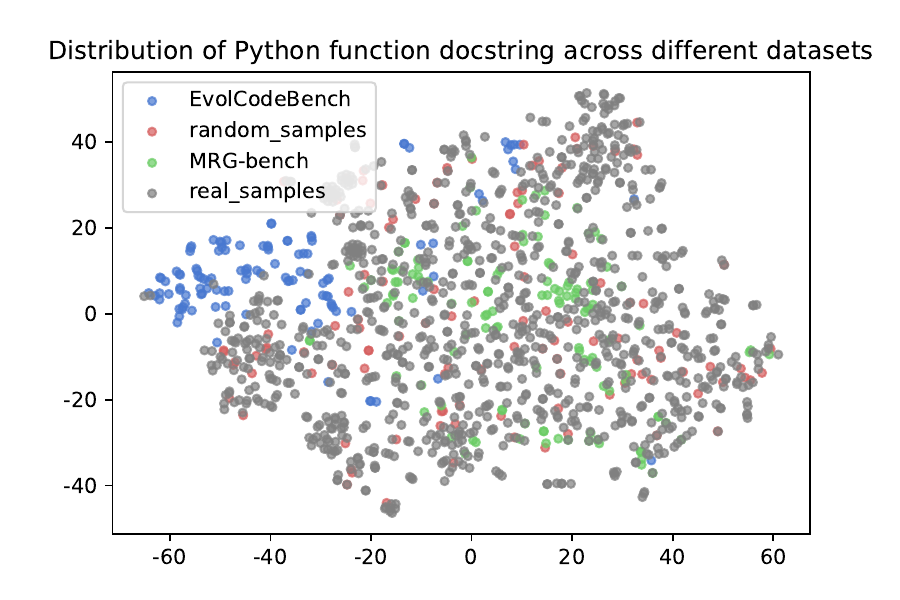}
        \caption{Distribution of samples for different Python datasets}
        \label{fig:subfig_a}
    \end{subfigure}
    
    \begin{subfigure}[b]{0.75\linewidth}
       \includegraphics[width=\linewidth]{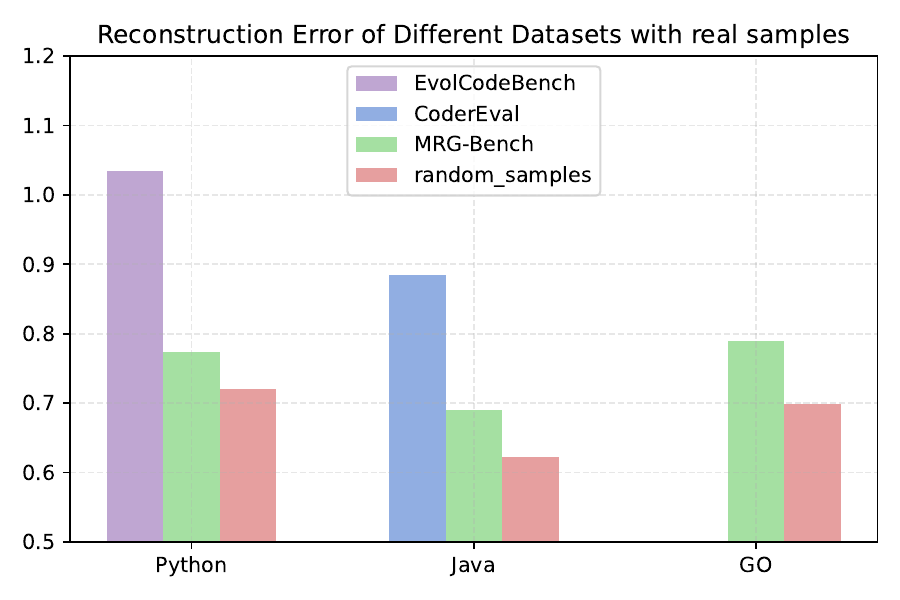}
        \caption{Reconstruction Error between different datasets and real data on 3 languages}
        \label{fig:subfig_b}
    \end{subfigure}
\end{figure}

\subsection{RQ1: Why do we need MRG-Bench?}

\subsubsection{RQ1.1: Is MRG-Bench more representative compared to existing repository-level code generation datasets?}

One of the most critical design principles of MRG-Bench is \textbf{Practicality}, as we aim for MRG-Bench to align closely with the actual distribution of open-source code. While some existing works have attempted to construct repository-level code generation datasets, only two datasets, EvoCodeBench~\cite{evobench} and AgentBench~\cite{xcodeeval}, provide complete repository information with executable unit tests. However, as the source code for AgentBench is unavailable, we compare the data from EvolBench with MRG-Bench in this section.

Specifically, we selected 500 high-quality, popular open-source projects (excluding the 22 projects in MRG-Bench) in Python, Java, and Go to create a representative dataset. From these projects, we randomly sampled 10,000 functions, extracting their docstrings and function bodies to represent "real-world projects." To effectively analyze the semantic distribution, we employed NV-Embedding-2~\cite{nvembed2}, one of the state-of-the-art embedding models, to embed each docstring into a 4,096-dimensional semantic vector. 

For comparison, we first established a random control group, referred to as the \textit{random-sampled dataset}, for each language by randomly extracting a dataset of equivalent size from open-source repositories. Additionally, we selected queries from existing repository-level code generation datasets for comparison. For Python, we chose \textit{EvolCodeBench} \cite{evobench}, and for Java, we selected data from the \textit{CoderEval-Java} \cite{codereval} subset. Although \textit{CoderEval} was not specifically designed as a repository-level code generation dataset, its data is indeed sourced from real open-source repositories. Unfortunately, there is currently no available dataset for the Go language, so we could only compare it with the randomly sampled dataset. We measure the gap between each dataset and the real query distribution (10,000 samples) using the \textbf{Reconstruction Error} of the sampled data in reconstructing the real distribution, \textbf{Reconstruction Error} is defined as follows:

Reconstruction of \( B \) using \( A \): For each \( b \in B \), find its nearest neighbor \( a \in A \) then compute the reconstruction error, \textbf{smaller is better}:

   \[
   \text{Reconstruction Error} = \frac{1}{|B|} \sum_{b \in B} \min_{a \in A} \|a - b\|_2
   \]

\( |A| \) and \( |B| \) represent the sizes (number of elements) of sets \( A \) and \( B \), respectively. \( \|\cdot\|_2 \) denotes the Euclidean norm.  
A smaller reconstruction error indicates that the sampled data is closer to the real data distribution.

Figure~\ref{fig:subfig_b} shows the reconstruction errors between different datasets to real-world data. We can find that MRG-Bench is significantly closer to the real-world distribution compared to EvoCodeBench and CoderEval-Java. The reconstruction error values of MRG-Bench are very close to those of random sampling. These results demonstrate that MRG-Bench represents the real-world code distribution more accurately. Consequently, the results obtained on this dataset better reflect the real performance of models.

To more clearly illustrate the distribution of the MRG-Bench, we used t-SNE to reduce the dimensionality of the data and visualized the data distribution for Python as an example, as shown in Figure~\ref{fig:subfig_a}. The gray data points in the figure represent real data, and for clarity of the plot, we further sampled 2,000 data points from the 10,000 total samples for visualization. We can find that the queries from EvolCodeBench (in blue) deviate significantly from the semantic distribution of real queries, with many semantics left uncovered. In contrast, MRG-Bench aligns more closely with the real distribution and the results of random sampling.

\textit{\textbf{Takeaway-1}: MRG-Bench demonstrates a significantly lower reconstruction error compared to other datasets, indicating its superior ability to represent real-world code semantics. Visualizations using t-SNE further confirm that MRG-Bench closely mirrors the semantic distribution of actual code, making it a more reliable benchmark for evaluating model performance in practical scenarios.}

\begin{table*}
\centering
\caption{Pass@1 and Pass@3 for different models on MRG-Bench.}
\label{tab:straight}
\begin{tabular}{lccccccccc}
\hline
\textbf{Model}   & \multicolumn{2}{c}{\textbf{Python}} & \multicolumn{2}{c}{\textbf{Java}} & \multicolumn{2}{c}{\textbf{Go}} & \multicolumn{2}{c}{\textbf{Average}} \\ 
                 & \textbf{pass@1} & \textbf{pass@3}   & \textbf{pass@1} & \textbf{pass@3}   & \textbf{pass@1} & \textbf{pass@3}   &  \textbf{pass@1} & \textbf{pass@3} \\ 
\hline
CodeLLaMA-13B    & 9.8\%  & 14.6\%  & 7.8\%  & 10.4\%   & 4.5\%  & 6.1\%   & 7.4\% & 10.4\%   \\ 
DeepSeek-Coder-33B         & 7.9\%  & 10.6\%  & 5.9\%  & 8.3\%   & 5.3\%  & 7.4\%   & 6.4\% & 8.8\%   \\ 
StarChat2-15B        & 7.9\%  & 10.6\%  & 5.9\%  & 9.4\%   & 5.3\%  & 7.4\%   & 6.4\% & 9.1\%   \\ 
LLaMA-3.1-8B-Instruct           & 7.3\%  & 8.9\%   & 5.6\%  & 8.9\%   & 1.2\%  & 1.2\%   & 4.7\% & 6.3\% &  \\ 
GPT-4o          & 8.7\%  & 10.6\%  & 5.6\%  & 8.9\%   & 4.9\%  & 4.9\%   & 6.4\% & 8.1\%   \\ 
Claude3.5-Sonnet          & \textbf{10.6\%} & \textbf{15.9\%}   & \textbf{7.8\%}  & \textbf{10.4\%}  & \textbf{7.3\%}  & \textbf{11.6\%}  & \textbf{8.6\%} & \textbf{12.6\%}   \\ 
\hline
\end{tabular}
\end{table*}

\subsubsection{RQ1.2 Why do we need a multi-language evaluation dataset?}

To demonstrate the multi-language advantages of MRG-Bench, we conducted experiments on the performance of LLMs across different programming languages. Specifically, we provided the models with the annotations and signatures of functions and asked them to complete the specified function. To verify the baseline capability differences of the models across languages, we did not provide any additional contextual information beyond the function details. The experimental results are shown in Table~\ref{tab:straight}. According to the results, it is evident that every model performs significantly better on Python compared to Java and Go. Moreover, Go, being a relatively niche language, exhibits the poorest performance among the three languages. For example, LLaMA-3.1-8B-Instruct achieves 7.3\% Pass@1 on Python but 1.2\% Pass@1 on GO.  This indicates that existing models exhibit significant bias toward different programming languages. Even Java, a relatively mainstream language, shows notably weaker performance compared to Python. These findings cannot be revealed by current evaluation datasets, which are predominantly focused on Python, which presents a challenge to previous evaluation efforts, as most earlier work focused on single-language evaluations, often concentrated on Python. This imbalance may lead to an overestimation of large language models' performance in other programming languages. The finding highlights the value of MRG-Bench's multi-language support—not only for evaluating performance on specific languages, but also for uncovering significant performance disparities of models across different programming languages.

\textit{\textbf{Takeaway-2}: MRG-Bench’s multi-language evaluation reveals significant performance disparities in LLMs across programming languages, illustrating how current benchmarks (often Python-centric) fail to capture these gaps.}

\subsection{RQ2: How do current LLMs perform on MRG-Bench with different contexts?}

\begin{table*}
\centering
\caption{Pass@1 and Pass@3 Providing In-file Context to Models.}
\label{tab:infile}
\begin{tabular}{lcccccccc}
\hline
\textbf{Model} & \multicolumn{2}{c}{\textbf{Python}} & \multicolumn{2}{c}{\textbf{Java}} & \multicolumn{2}{c}{\textbf{Go}} & \multicolumn{2}{c}{\textbf{Average}} \\
 & \textbf{Pass@1} & \textbf{Pass@3} & \textbf{Pass@1} & \textbf{Pass@3} & \textbf{Pass@1} & \textbf{Pass@3} & \textbf{Pass@1} & \textbf{Pass@3} \\ 
\hline
CodeLLaMA-13B & 13.6\% & 24.4\% & 12.3\% & 21.5\% & 13.9\% & 23.9\% & 13.3\% & 23.3\% \\
DeepSeek-Coder-33B & 25.5\% & 34.1\% & 21.5\% & 26.1\% & 19.6\% & 29.4\% & 22.2\% & 29.9\% \\
LLaMA-3.1-8B-Instruct & 18.1\% & 22.8\% & 16.9\% & 20.0\% & 13.5\% & 20.8\% & 16.2\% & 21.2\% \\
GPT-4o & 33.3\% & 38.2\% & 24.6\% & 29.2\% & 28.8\% & 33.7\% & 28.9\% & 33.7\% \\
Claude3.5-Sonnet & 33.3\% & 41.5\% & 29.2\% & 40.0\% & 35.0\% & 47.2\% & 32.5\% & 42.9\% \\
\hline
\end{tabular}
\end{table*}

\subsubsection{RQ2.1: How do different models perform when provided with fixed context?}
In this experiment, we employ two types of fixed contextual information: (1) Infile context: The source code containing the target function to be generated. (2) Callee Context: The callee function information of the target function, efficiently extracted through our function call analysis framework.  These two information respectively represent the context surrounding the target function and available custom function definitions that may be invoked. In addition to contextual settings, we also assessed the performance of state-of-the-art models on MRG-Bench to understand their capabilities in repository-level code generation.

We first investigated how much improvement can be achieved by providing the file content in which the function is located, as this information is the easiest to obtain in practice. Specifically, we prepended the file content to the beginning of the function generation prompt and asked the model to generate the function based on the given annotations and function signature. Since different models use different tokenizers, we used the tokenizer from GPT-3.5-Turbo to truncate the file content to a maximum of 6000 tokens, ensuring that no model exceeded its maximum token length. Due to the shorter context window of StarChat2-15B, this model was excluded from this experiment. The results are shown in the Table \ref{tab:infile}.

Based on the results, we find that providing contextual information about the function significantly improves the Pass@1 rate. For example, the Pass@1 of CodeLLaMA-13B shows noticeable improvement from an average of 7.4\% to 13.3\%. After incorporating relevant context, DeepSeek-Coder-33B demonstrates a substantial advantage, outperforming CodeLLaMA-13B and LLaMA-3.1-8B. Moreover, DeepSeek-Coder-33B exhibits strong competitiveness with closed-source models; for instance, its Pass@3 score is comparable to the pass@1 performance of GPT-4o and Claude3.5-Sonnet.

Additionally, we find that while the performance gap between languages was substantial in RQ1.2, the differences between models across different languages narrowed significantly after providing rich context.  This implies that evaluation outcomes may vary significantly across programming languages under different contextual conditions, necessitating an expansion of current assessment methodologies.

\textit{\textbf{Takeaway-3}: Providing in-file contextual information significantly improves the performance of large language models (LLMs). Additionally, the performance gap between languages narrowed when context was included.  This implies that evaluation outcomes may vary significantly across programming languages under different contextual conditions, necessitating an expansion of current assessment methodologies.}

\begin{table*}
\centering
\caption{Performance Comparison with Different Contexts}
\label{tab:callee}
\begin{tabular}{lcccccccc}
\hline
\textbf{Different Context} & \multicolumn{2}{c}{\textbf{Python}} & \multicolumn{2}{c}{\textbf{Java}} & \multicolumn{2}{c}{\textbf{Go}} & \multicolumn{2}{c}{\textbf{Average}} \\
 & \textbf{Pass@1} & \textbf{Pass@3} & \textbf{Pass@1} & \textbf{Pass@3} & \textbf{Pass@1} & \textbf{Pass@3} & \textbf{Pass@1} & \textbf{Pass@3} \\ 
\hline
callee-funcbody & 19.60\% & 22.80\% & 11.30\% & 15.10\% & 7.50\% & 10.60\% & 12.8\% & 16.2\% \\
callee-signature & 13.30\% & 15.80\% & 7.60\% & 10.40\% & 5.10\% & 8.30\% & 8.7\% & 11.5\% \\
in-file & 25.50\% & 34.10\% & 21.50\% & 26.10\% & 19.60\% & 29.40\% & 22.2\% & 29.9\% \\
\hline
\end{tabular}
\end{table*}

A major challenge in repository-based code generation is that the model lacks knowledge of the available functions within the repository. Thanks to our analysis framework, we are able to extract the repository functions utilized by the target function during the generation process and provide them to the model as context. In this section, we conducted experiments on the previously identified optimal open-source model, DeepSeek-Coder-33B~\cite{deepseekcoder}. All the callable functions are concatenated into the prompt and provided as input to the model, which is tasked with generating the specified function based on the given annotations and function signature. The experimental results are shown in Table \ref{tab:callee}.

Based on the data in the table, we can observe that while providing available functions brings some improvement to the performance (DeepSeek-Coder-33B achieves a Pass@1 score of 19.62\% on Python), these improvements are far smaller than those brought by providing file content. This indicates that when generating the code we need, the model is not simply encountering difficulties in \textbf{how to accomplish} the expected functionality (since we have already provided all custom functions). This motivates us to further investigate the model's erroneous results to explore what repository context the model truly needs during generation.

\textit{\textbf{Takeaway-4:} Providing available functions yields less improvement than providing in-file context, suggesting that the model's difficulties lie not simply in \textbf{how to accomplish} the expected functionality, but rather in other underlying issues.}

\begin{table}
\centering
\caption{Pass@1 Performance Comparison of Different Methods and Models}
\label{tab:longr}
\begin{tabular}{llccc}
\hline
\textbf{Method} & \textbf{Model} & \textbf{Python} & \textbf{Java} & \textbf{Go} \\ 
\hline
In-file Context & Claude3.5-Sonnet & 33.3\% & 29.2\% & 35.0\% \\
\hline
\multirow{2}{*}{Long Context} 
 & Claude3.5-Sonnet & 31.7\% & 33.9\% & 33.7\% \\
 & DeepSeek-V2.5 & 25.2\% & 32.3\% & 28.8\% \\
\hline
\multirow{3}{*}{Reasoning Model}
 & Deepseek-R1 & \textbf{34.15}\% & 33.33\% & \textbf{39.26}\% \\
 & O3-mini & \textbf{34.15}\% & \textbf{38.54}\% & 32.52\% \\
 & QWQ-32B & 13.01\% & 15.62\% & 16.56\% \\
\hline
\end{tabular}
\end{table}

In order to further explore the optimal performance of the current models, we conducted experiments on longer context window and stronger reasoning models.  For long context models, we tested the optimal open-source model, DeepSeek-V3, and the optimal closed-source model, Claude3.5-Sonnet. DeepSeek's official API supports a 128K token context window. Therefore, we concatenated all code from the \textbf{folder} containing the target function into a single context, which was then input into the model for generation. To ensure consistency between the two models, we standardized the context size to 100K tokens, truncating any data exceeding this length. The results for both models are presented in Table \ref{tab:longr}. For reasoning models, we select O3-mini, Deepseek-R1~\cite{deepseekr1} and Qwen2.5-QWQ-32B~\cite{qwq}. We use the same in-file context in RQ2.1 for the experiments of reasoning models. The results are shown in Table~\ref{tab:longr}.

As shown in the data, while the performance of long-context models surpasses that of In-file-context method, the improvement is not substantial. For instance, Claude3.5-Sonnet get worse performance on Python and Go with longer context. This suggests that, although longer contexts can provide more useful information, they may also introduce additional noise. A more promising approach may be to selectively provide different contexts tailored to the requirements of each programming language.

Furthermore, our experiments reveals that reasoning-enhanced models, when combined with in-file contextual information, demonstrate superior performance. Specifically, DeepSeek-R1 and O3-mini achieve state-of-the-art results on MRG-Bench. However, even the most advanced large language models (LLMs) currently available exhibit a Pass@1 accuracy below 40\%, underscoring the significant challenge MRG-Bench poses to current LLMs.

Notably, QWQ-32B underperforms compared to DeepSeek-Coder-33B, despite sharing the same parameter scale. We hypothesize that this performance gap may stem from QWQ-32B’s reasoning-oriented training, which could compromise its foundational code generation capabilities.

\textit{\textbf{Takeaway-5:} While long-text models perform better than the In-file-context method on MRG-Bench, the improvements are not substantial. The reasoning models achieved the highest performance, yet none surpassed 40\% in Pass@1 accuracy, demonstrating that MRG-Bench remains a highly challenging benchmark for state-of-the-art models.}

\subsubsection{RQ2.2: How do retrieval-augmented generation (RAG) approaches perform on MRG-Bench when supplying dynamically retrieved context?}

\begin{table*}
\centering
\caption{Pass@1 and Pass@3 of Different RAG Methods Using DeepSeek-Coder-33B on MRG-Bench}
\label{tab:rag_results}
\begin{tabular}{lcccccccc}
\hline
\textbf{RAG Method} & \multicolumn{2}{c}{\textbf{Python}} & \multicolumn{2}{c}{\textbf{Java}} & \multicolumn{2}{c}{\textbf{Go}} & \multicolumn{2}{c}{\textbf{Average}} \\
 & \textbf{Pass@1} & \textbf{Pass@3} & \textbf{Pass@1} & \textbf{Pass@3} & \textbf{Pass@1} & \textbf{Pass@3} & \textbf{Pass@1} & \textbf{Pass@3} \\ 
\hline
mix-rag & 15.80\% & 18.40\% & 11.30\% & 15.10\% & 9.80\% & 14.20\% & 12.3\% & 15.9\% \\
bm25-rag & 13.30\% & 17.10\% & 9.40\% & 14.20\% & 7.50\% & 10.60\% & 10.1\% & 14.0\% \\
bge3-rag & 17.10\% & 19.00\% & 11.30\% & 16.00\% & 10.20\% & 15.00\% & 12.9\% & 16.7\% \\
repocoder & 15.80\% & 18.40\% & 20.80\% & 21.70\% & 17.70\% & 20.10\% & 18.1\% & 20.1\% \\
\hline
\end{tabular}
\end{table*}

Repository-level code generation can be categorized as a Retrieval-Augmented Generation (RAG) task, wherein the model retrieves relevant information to support the generation process.  To evaluate the performance of various methods on MRG-Bench, we implemented several basic RAG techniques. Specifically, we designed four sets of experiments: (1) the RAG method using BM25 retriever, (2) the RAG method using embedding-based retriever, (3) a mixed RAG method combining both above retrieval techniques, and (4) Repocoder, a code generation method designed specifically for repository-level tasks. For all methods, we employed DeepSeek-Coder-33B as the generation model, while the BGE-M3 model, recognized for its strong performance, was used for embeddings the code snippet. We adopted the code document block segmentation strategy provided by LangChain, segmenting the codebase into blocks within 500 tokens with an overlap of 50 tokens. During retrieval, we selected the top 5 most relevant code blocks as the context for generation. For Repocoder, we used the hyperparameters specified in the paper. The experimental results are presented in the Table \ref{tab:rag_results}.

The results reveal a substantial performance disparity between RAG-based models across different programming languages. Among the three languages evaluated, Python achieved the highest performance, while Go performed the worst. For instance, in the mix-RAG method, the Pass@1 score of Python was double that of Go. Moreover, when using BM25 as a retriever, pass@1 scores for all languages were lower compared to using embedding-based retrieval. After excluding the BM25 retriever, the performance of the embedding-only RAG method surpassed that of the mixed retrieval approach, indicating that BM25 can be replaced by embedding-based methods without sacrificing performance.

Repocoder, designed for repository-level code generation, did not outperform the basic RAG method for Python but demonstrated superior performance in Java and Go, significantly surpassing the RAG baseline. Notably, Repocoder showed consistent performance across all three languages, suggesting it does not exhibit language bias. We attribute this balanced performance to Repocoder's use of pre-generated code for retrieve, ensuring that both the query and retrieved information reside in the same semantic space, thus improving retrieval accuracy. In contrast, other methods suffer from inadequate retrieve information, leading to similar issues as observed in RQ1 and RQ2, with Python outperforming other languages.

\textit{\textbf{Takeaway-6:} Although RAG-related methods achieve certain performance improvements, their effectiveness is significantly inferior to that of providing in-file context alone. This further suggests that the primary challenges faced by current methods and the key contextual information require further exploration. Notably, RepoCoder demonstrates exceptional capability in eliminating language bias, which warrants deeper investigation.}

\begin{figure}
    \centering
    \includegraphics[width=1\linewidth]{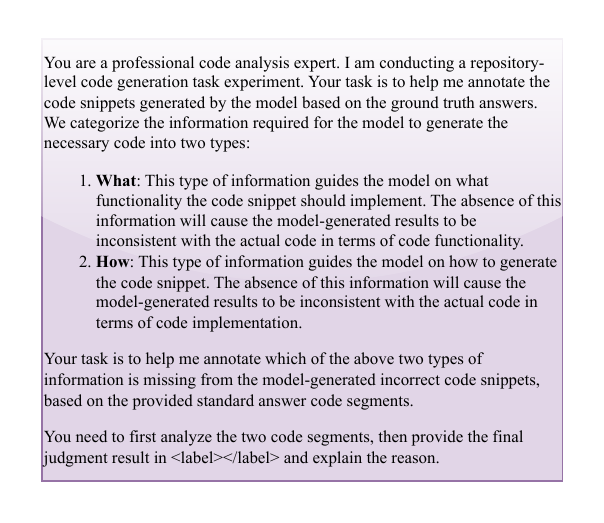}
    \caption{Prompt for label fail cases to How and What.}
    \label{fig:prompt}
\end{figure}

\begin{figure}
    \centering
    \includegraphics[width=1\linewidth]{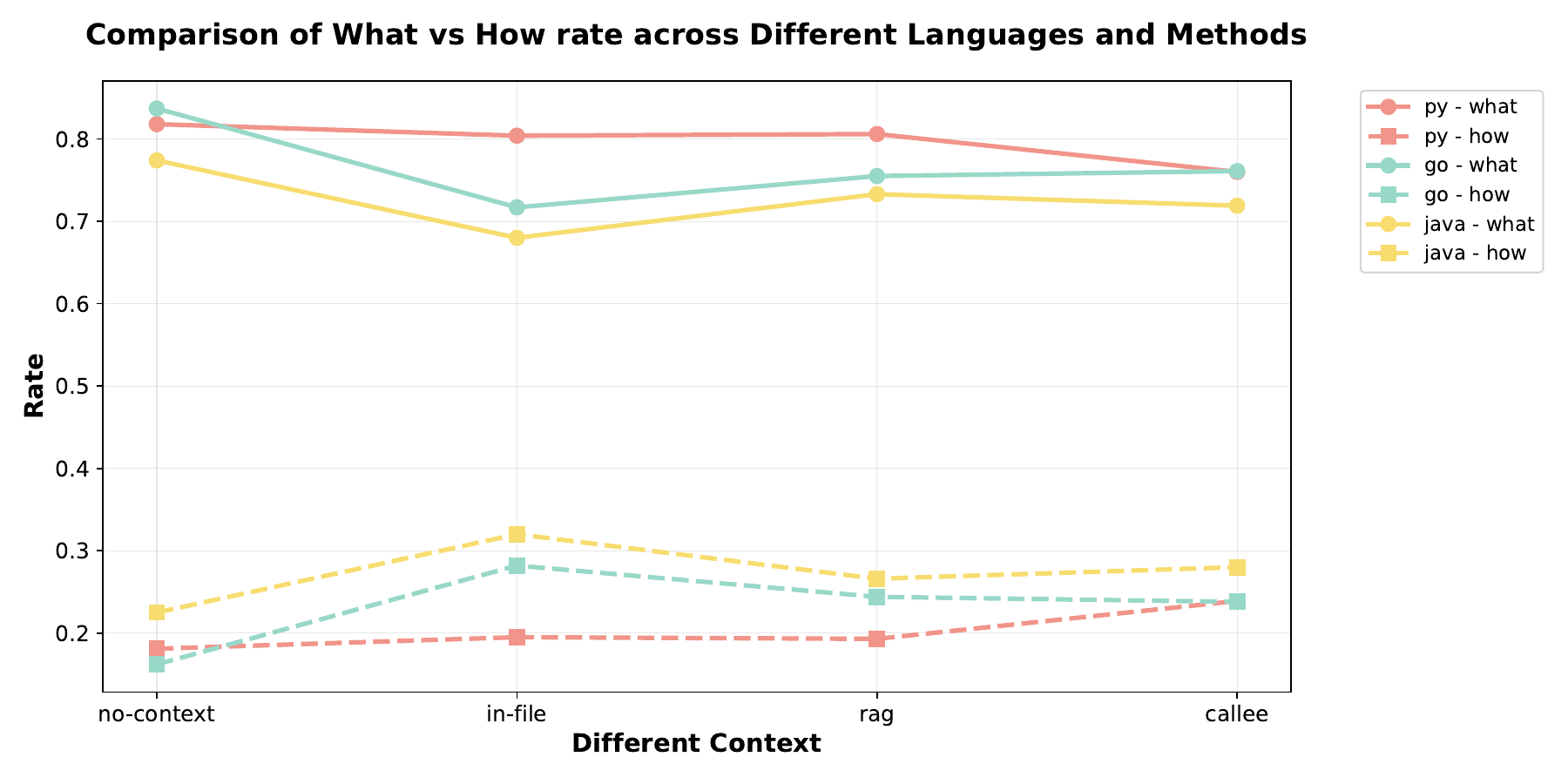}
    \caption{Annotation result of fail reason.}
    \label{fig:whathow}
\end{figure}

\subsection{RQ3:Why do current methods fails on MRG-Bench?}

From the previous two RQs, we observe several intriguing phenomena: on MRG-Bench, RAG algorithms show no significant advantages, and providing available functions for code generation offers minimal benefit to models. Instead, simply providing the file containing the target code proves more effective. To explain these phenomena, we attempt to annotate the causes of model failures, offering novel insights into the contextual requirements of models. To our knowledge, this is the first research effort to explore the relationship between model failure causes and context types.

Inspired by Takeaway-4 (RQ2.1), we found that providing custom functions needed to "implement target functionality" yields limited model improvement, suggesting that generation difficulties may lie in other perspectives. Following intuitive problem-solving logic, we categorize the information required for models to generate specified code into two types: (1) \textbf{What to do} - information that guides models to understand the functional logic required in the target function, potentially including function definitions, comments, classes, sibling functions, or functions with similar names. We call this "\textbf{What information}". (2) \textbf{How to do} - information that guides models on implementation approaches, such as appropriate frameworks or callable custom functions. we call this "\textbf{How information}". To understand which type of missing information causes model failures, we annotate failure cases from the currently best-performing model, Claude-3.5-Sonnet.

We employ a carefully designed prompt (shown in Figure~\ref{fig:prompt}) and utilize five top-performing models (GPT-4o, Claude-3.5-Sonnet, Gemini-2.5-Pro, DeepSeek-V3, and Qwen2.5-72B-Instruct) for annotation and voting to produce final results, requiring models to identify failure causes in unsuccessful cases. To maintain annotation stability, all generation processes use temperature=0. To further ensure annotation reliability, we only retain samples with voting ratios of 5:0 and 4:1, discarding highly controversial 3:2 cases. The final retention rate is 86.3\%, with most cases receiving consistent annotations across models.

Our annotation results are shown in Figure~\ref{fig:whathow}:

The results reveal commonalities across different languages. First, over 68\% of failures stem from missing "What information" - models cannot understand what code functionality the user requirements correspond to within the current repository context. Providing different contextual information from repositories can effectively reduce this proportion, helping models understand the working environment and make more precise judgments. However, these improvements are quite limited (infile, RAG), suggesting we should further enhance retrieval techniques for this type of information, such as repository README documents, feature descriptions, and specific application scenarios.

\textit{\textbf{Takeaway-7}: Missing "What information" is the primary cause of current model failures; models cannot accurately understand what functionality requirement descriptions correspond to within the specific repository context.}

Benefiting from MRG-Bench's multilingual feature, we find that different programming languages exhibit varying contextual preferences. Python is particularly notable (recall that Python was also special in evaluation of RQ2). Python demonstrates a distinctly high "What Information" tendency. Across different contexts (except No Context), Python's lowest "What" proportion equals the maximum values of the other two languages. This indicates that in Python, once models struggle to understand specified requirements, repository context rarely provides useful information. Combined with conclusions from RQ1.2, this yields an interesting inference: Python language itself exhibits weaker semantic associations between functions compared to other languages. Therefore, even without context, models can achieve good performance (Takeaway-2, RQ1.2). If models fail, providing context rarely resolves the issue. Only providing truly needed sub-functionalities (callable functions) for target functions proves effective, but such context is extremely difficult to obtain in real applications.

When attempting to improve model performance in practical scenarios, for Go and Java, we can follow in-file and RAG approaches to further mine "What Information" while eliminating ineffective information, expecting better results. However, for Python, these approaches are ineffective. We need to start from RepoCoder's approach, attempting to use fine-grained information from target functions to reversely summarize "What information," presenting new challenges for further improving practical application effectiveness.

\textit{\textbf{Takeaway-8}: Python exhibits markedly different contextual preferences compared to the other two languages. Following RepoCoder's pre-generation approach to further mine "What Information" represents a potentially effective direction. Different languages in real application scenarios may require different context extraction strategies.}

\section{Threats to Validity}
\label{sec:threats}
Our work faces the following threats:

\textbf{Low Coverage of Languages and Projects.} Although we initially screened 7 languages and 1,400 projects, only 3 languages and 22 projects were retained in the final dataset. This may lead to a variance in our evaluation results.  However, our experiments (RQ1.1) demonstrate that MRG-Bench, despite encompassing fewer projects, exhibits a closer alignment with real-world data distributions compared to alternative datasets. In practice, contemporary high-quality open-source projects typically comprise complex, multi-module systems containing code segments with diverse functionalities. Our streamlined selection strategy—which neither imposes restrictions on function dependencies nor modifies the original project structure during test environment construction—enables maximal preservation of source code distributions.  In contrast, other datasets incorporating more projects often introduce distributional biases during their construction processes due to sophisticated sampling criteria and artificial dependency pruning.  Even though, we plan to expand MRG-Bench's project coverage in future iterations to cover more functionalities, and expand language of C/C++ and JavaScript by building isolate docker environment for each project.

\textbf{Lack of agent based method.} We did not include agent-related methods in the paper because these often require extensive prompt configuration and involve network search functions, which may cause data leakage and affect the fairness of the evaluation. Besides, current mainstream open-source agent methods are primarily designed for bug-fixing scenarios in SWE-Bench, making them difficult to adapt to the generation scenarios in MRG-Bench. We look forward to the emergence of agent methods specifically designed for general code generation, and will update the leaderboard with relevant approaches when they become available.

\begin{table}[h]
\centering
\caption{Data Leakage Detection Results}
\label{tab:dataleakage}
\begin{tabular}{l|cc}
\hline
\textbf{Dataset} & \textbf{Model} & \textbf{Data Leakage Ratio(CDD)} \\ 
\hline
HumanEval & GPT-4o & 41.47\% \\
\hline
\multirow{6}{*}{MRG-Bench} 
& GPT-4o & 7.2\% \\
& Claude3.5-Sonnet & 7.8\% \\
& DeepSeek-Coder-33B & 6.0\% \\
& CodeLLaMA-13B & 4.9\% \\
& LLaMA-3.1-8B-Instruct & 7.7\% \\
\hline
\end{tabular}
\end{table}
\textbf{Data Leakage Problem.} Data leakage of dataset is one of the most critical issues in evaluating large language models. We address this problem from two aspects. First, we select newer repositories that have less relevant information available online, are less imitated and cited by other repositories, and with which the models are less familiar. Second, we use data leakage detection approach CDD~\cite{cdd} to check the dataset leakage ratio (Using default parameters in their paper). CDD can detect whether LLMs have been
trained on specific benchmarks by compare the generated result in different sampling tempreture. The detection results are shown in
Table~\ref{tab:dataleakage}. Compared to HumanEval~\cite{humaneval}, the leakage ratio on MRG-Bench is low.
Given that the generated outputs share identical function signatures, documentation, and partial reuse of public APIs/code elements, it is inherently challenging to reduce this metric to zero. Based on the information above and the suboptimal performance of current models on MRG-Bench, we can conclude that the benchmark exhibits low susceptibility to data leakage.

\section{Related Work}
\subsection{Large Language Models for Code Generation}

The advent of pre-training technology has significantly propelled the field of code generation, leading to remarkable advancements in both academia and industry \cite{Shen2022, Nijkamp2023, Fried2023}. This surge has resulted in the emergence of various large language models (LLMs) that have demonstrated excellent performance in code generation tasks. Notable examples include Codex \cite{codex}, ChatGPT \cite{chatgpt}, CodeLLaMA \cite{codellama}, DeepSeek Coder \cite{deepseekcoder}, and StarCoder2 \cite{starcoder2}.

\subsection{Code Generation Benchmarks}

Early benchmarks for code generation primarily evaluated LLMs on generating relatively simple Python functions. HumanEval \cite{humaneval} and MBPP \cite{mbpp} consist of manually designed programming questions that provide function signatures, comments, and unit tests. LLMs are tasked with crafting function bodies based on these inputs, and their outputs are evaluated by the success in passing the provided unit tests. However, these benchmarks focus on self-contained functions with only built-in language dependencies, which do not fully represent real-world development scenarios \cite{codereval}.

To address these limitations, more complex benchmarks have been developed. APPS \cite{apps} evaluates code generation on more challenging competition-style problems.  CoderEval \cite{codereval} introduce non-standalone programs derived from real GitHub projects, aligning better with actual development settings that rely on multiple public libraries and project files.

Multi-task benchmarks like CodeXGLUE \cite{codexglue} and XCodeEval \cite{xcodeeval} incorporate a broad range of programming questions and tasks, establishing a comprehensive framework for evaluating LLMs. However, CodeXGLUE relies on textual similarity metrics such as BLEU and CodeBLEU \cite{codescore}, which may not fully capture the functional correctness of code. CoderUJB \cite{coderujb} fills a critical gap by providing a benchmark that includes multiple programming tasks that are executable and match real-world development scenarios. 

Recently, benchmarks for repository-level tasks have been proposed. CrossCodeEval \cite{crosscodeeval}, RepoBench \cite{repobench}, and RepoEval \cite{repocoder} are code completion benchmarks that aim to evaluate LLMs on code completion tasks within repositories. However, they lack necessary runnable environment, limiting their applicability for code generation evaluation. SWE-bench \cite{swebench} focuses on repairing repository issues by revising existing programs rather than generating new code.

However, existing datasets present several issues, including inconsistencies with pratical code information, a lack of runnable test cases, and limited language coverage. To address these shortcomings, we propose \textbf{MRG-Bench}. Our dataset encompasses three programming languages and exclusively includes data from high-quality open-source projects. This ensures that \textbf{MRG-Bench} provides an effective evaluation framework for subsequent repository-oriented code generation tasks.

\section{Conclusion and Future Work}

In this paper, we present a code generation evaluation dataset based on an analysis of high-quality open-source code repositories. Our dataset exhibits three key characteristics: \textbf{Practicality:} The dataset is derived from real-world open-source projects, avoiding the use of artificially constructed data. \textbf{Multilingualism:} The dataset includes a diverse range of programming languages to enable comprehensive evaluations across different languages.\textbf{Executability:} The primary evaluation criterion for the dataset is that the code generated by models should be functional in real-world scenarios.

In addition, we conducted a comprehensive evaluation of current large language models and Retrieval-Augmented Generation (RAG) methods using MRG-Bench. Our experimental results indicate that the accuracy of existing large language models in solving real-world code generation tasks remains low. While providing context can improve performance, the highest Pass@1 score does not exceed 40\%. This highlights the need for further attention to evaluating large language models in real application scenarios. 

Finally, we designed a novel annotation methodology to investigate the underlying causes of current method failures. Experimental results demonstrate that the primary challenge for current approaches stems from "\textbf{understanding user input, i.e., determining what to do.}" The repository context that enhances this capability exhibits distinct preferences across different programming languages, suggesting the need for more diverse context extraction techniques. Notably, Python demonstrates markedly different characteristics from other languages in both performance and contextual preferences. Therefore, we encourage future research to expand the scope of programming languages to further explore the characteristics of both models and languages.

All our data, code, experiment results(generated code and test log) are public available on Github\footnote{https://github.com/MRG-Bench/MRG-Bench}


\bibliographystyle{ACM-Reference-Format}
\bibliography{sample-sigconf-authordraft}

\end{document}